\documentclass[conference]{IEEEtran}
\usepackage[utf8]{inputenc}
\usepackage{cite}
\usepackage{color}
\usepackage{textgreek}
\usepackage{graphicx}
\usepackage{flushend}

\begin{document}

\title{Emerging Biometric Modalities and their Use: Loopholes in the Terminology of the GDPR and Resulting Privacy Risks}

\author{\IEEEauthorblockN{Tamas Bisztray\IEEEauthorrefmark{1}, Nils Gruschka\IEEEauthorrefmark{1}, Thirimachos Bourlai\IEEEauthorrefmark{2}, Lothar Fritsch\IEEEauthorrefmark{3}}

\IEEEauthorblockA{\IEEEauthorrefmark{1}Department of Informatics, University of Oslo, Oslo, Norway\\
Email: \texttt{\{tamasbi,nilsgrus\}@ifi.uio.no}}
\IEEEauthorblockA{\IEEEauthorrefmark{2}University of Georgia, Athens, Georgia, United States\\
Email: \texttt{thirimachos.bourlai@uga.edu}}
\IEEEauthorblockA{\IEEEauthorrefmark{3}Oslo Metropolitan University, Oslo, Norway\\
Email: \texttt{lothar.fritsch@oslomet.no}}
}


\maketitle

\begin{abstract}
Technological advancements allow biometric applications to be more omnipresent than in any other time before. This paper argues that in the current EU data protection regulation, classification applications using biometric data receive less protection compared to biometric recognition. We analyse preconditions in the regulatory language and explore how this has the potential to be the source of unique privacy risks for processing operations classifying individuals based on soft traits like emotions. This can have high impact on personal freedoms and human rights and therefore, should be subject to data protection impact assessment.
\end{abstract}


\section{Introduction}

The General Data Protection Regulation (GDPR) \cite{GDPR} is the data protection regulation of the European Union (incl. the European Economic Area). All processing of personal data of EU citizens must comply with this regulation. In particular the processing of biometric data has the potential to impose high-risks to the rights and freedoms of individuals.
The exact definition of biometric data is not always consistent across legal and scientific publications or standards \cite{article_29_working_party_wp193_2012,meints_biometric_2008,jain_50_2016,4815263}. Thus, it can be difficult 
to interpret the resulting guidelines, implement appropriate privacy controls, or to conduct a data protection impact assessment (DPIA) as defined in Article 35 of the GDPR (which is required upon processing of biometric data according to \cite{kindt_having_2018}).

As per the GDPR, data defined as biometric does not receive any special protection when compared to other personal data such as name, email addresses, usernames, etc. Only when the processing of biometric data is done ``\textit{for the purpose of uniquely identifying a natural person}'', the data falls under the protection of Article 9. This article prohibits processing unless special conditions are fulfilled (e.g, the data subject has given explicit consent).
This means that other processing purposes using biometric data (unlike identification) only require the same level of protection as processing ``ordinary'' personal data. 
As an example, emotional reactions are measurable physiological processes and in a study researchers showed how seeing positive or negative social media posts can manipulate the emotions of people experimenting on 689,003 individuals without their knowledge \cite{kramer_experimental_2014}. In such situations the rights and freedoms of individuals can be threatened and yet, Article 9 would not apply as the purpose of processing is not identification. Secondly, whether such data can classify as biometric under the interpretation of the GDPR must be examined as Article 9 is only applicable to special categories not to personal data in general.
In this paper we analyse the definition of biometric data in the GDPR and compare it to other existing definitions. 
We show how preconditions in the GDPR are excluding towards certain biometric modalities and modes of operation namely, soft biometrics and classification. Finally, we will show risks resulting from the limited protection for these modes of operation.

The GDPR doesn't point to other standards such as the ISO/IEC 2382-37:2017(E) (ISO standard), to interpret the language it uses therefore, in the remainder of this paper the default interpretation of the terminology should be according to the taxonomy of the GDPR unless specified otherwise. Section 2 presents the Methodology for our analysis, Section 3 examines preconditions in the legal taxonomy, Section 4 discusses classification as a mode of operation, while Section 5 presents unique risks to rights and freedoms using soft biometrics for classification purposes. Section 6 summarises the results \footnote{
This is an updated version of DOI: 10.1109/BIOSIG52210.2021.9548298 and
includes language enhancements. The changes do not affect the
paper’s scope, analysis, and derived conclusions. The original version of this research is included in the proceedings of the 20th International Conference of the Biometrics Special Interest Group (BIOSIG). Editor: Tamas Bisztray. Date: October 2022 
}

\section{Methodology}
This paper utilises the WPR (\textit{What's the problem represented to be}) approach developed by Bacchi with the aim of policy analysis \cite{bacchi_why_2012}. It constitutes of the following seven steps: (i) WPR apprehends that policies can contain an implicit representation of the problem they are aiming to solve, by constructing a representation of reality which the policy responds to. (ii) This representation is based on (iii) presuppositions and assumptions that often (iv) omit and silence other aspects of reality, (v) and as such it produces a series of undesirable effects for the subjects in question. (vi) This requires the representation to be analysed and altered, leading to a new problem representation which then needs to be (vii) analysed again by the WPR approach. 

In our analysis this approach translates to: (i) examining how the GDPR defines and protects biometric data (ii) where the description of reality only considers recognition as the purpose of use, (iii) requiring the notions of ``specific technical processing'' and ``uniqueness'' (Art. 4 GDPR). (iv) Tying the definition and protection of data to preconditions, involuntarily silences classification purposes and the processing of non-unique biometric information. 
The main focus of this paper is to (v) highlight a series of undesirable effects stemming from these silences, in the form of privacy risks that are impacting rights and freedoms of natural persons.
Based on this the following (vi) problem representations will be examined: 
1. The inclusion of specific technical processing in the definition silences and excludes certain forms of biometric information from the category of biometric data.
2. Requiring that biometric data must allow or confirm unique identity can exclude forms of processed biometric information such as soft biometric templates.
3. Tying the protection provided by Article 9 to ``purpose of use'' is excluding towards other modes of operation such as verification and classification, based on biometric traits and characteristics of individuals.

We use discourse analysis as a methodology scrutinising legal text where our focus is to highlight risks to rights and freedoms of individuals. The arrangement of this paper and the presentation of the results of our analysis will follow a thematic line of thought instead of the repetitive steps of the WPR approach. Note, that in this paper our main focus is to outline the problem and the resulting risks, not to propose modifications to the GDPR.
Related work regarding legal analysis, privacy evaluation, and biometric technologies will be referenced at the relevant sections respectively. 
In \cite{kindt_having_2018} Kindt argued that the artificial distinction the GDPR makes between categories of biometric data based on ``specific technical processing'' and ``purpose of use'' is unnatural and calls for further discussion about this definition. 
In the following we highlight how these distinctions can translate to unique privacy risks.
We further expand the context of the discussion by the inclusion of the silenced notions of classification and soft biometrics.

\section{Taxonomy adapted by regulators and resulting risks}
Defining biometric data is not that simple as it seems on the first glance. One can find different definitions in relevant sources like legal regulations, technical standards and scientific literature. The term biometrics comes from the ancient Greek words `\textbeta\textiota\textomikron\textvarsigma' (bios) for life and `\textmu\textepsilon\texttau\textrho\textomikron\textnu' (metron) for measure, it is the measurement of traits and characteristics of living beings \cite{schatten_towards_2009}.
Most sources in literature acknowledge physical, physiological and behavioural traits, while others use categories such as biological, physiographic, motoric or biochemical \cite{bygrave_article_2020}.

The GDPR defines biometric data in Article 4.14 as: ``\textit{personal data resulting from `specific technical processing' relating to the physical, physiological or behavioural characteristics of a natural person, which `allow or confirm the unique identification' of that natural person, such as facial images or dactyloscopic data}''.  As shown in Fig. 1, this puts presuppositions on personal data containing measured traits and characteristics of human beings, to be recognised legally as biometric data. This together with preconditions in Article 9 can dampen the protection of such personal data. 
\begin{figure*}
    \centering
    \includegraphics[scale=0.91]{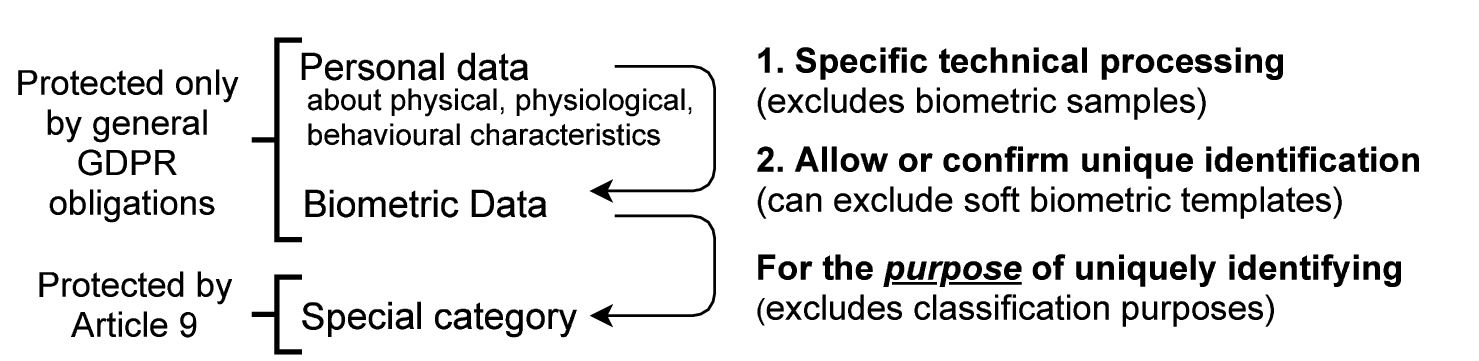}
    \caption{Conditions in the GDPR} 
    \label{fig:dimensions}
\end{figure*}

Personal data in question must relate to physical, physiological, and/or behavioural characteristics. This has to go through specific technical processing, where the resulting data allows or confirms identity, for such personal data to qualify as biometric. The order of the steps is not interchangeable. Article 9 only applies for data produced this way when its used for identification purposes. In contrast a proposed definition by the European Commission which was not adapted, defined biometric data as: ``\textit{Any personal data relating to the physical, physiological or behavioural characteristics of an individual which allow their unique identification, such as facial images, or, dactyloscopic data.}'' \cite{jasserand_avoiding_2015}. This definition doesn't require specific technical processing.

Analysing the notion ``allow or confirm the unique identification'' Jasserand argues that ``allow'' refers to establishing the identity (biometric \textit{identification}), whereas ``confirm'' refers to \textit{verifying} identity, where together these correspond to the notion of \textit{recognition} \cite{jasserand_avoiding_2015,jasserand_legal_2016}.
As Article 9 only mentions uniquely identifying, Bygrave and Tosoni concludes that verification purposes are excluded \cite{bygrave_article_2020}. 
Considering recognition as the only mode of operation is a shortcoming not unique to the GDPR, but shared by several legal articles, scientific papers and standards, aiming to give a technological overview, including the ISO standard.

Article 29 Working Party (WP29) defines biometric data as: ``\textit{biological properties, behavioural aspects, physiological characteristics, living traits or repeatable actions where those features and/or actions are both unique to that individual and measurable, even if the patterns used in practice to technically measure them involve a certain degree of probability.}'' 
This only requires the measurable features to be unique regardless of use, thus neither tying to ``specific technical processing'' nor to ``recognition''. However, requiring uniqueness can be silencing towards soft biometric traits. 
Soft biometric traits are defined by Dantcheva et al. as: ``\textit{physical, behavioural, or material accessories, which are associated with an individual, and which can be useful for recognising an individual. These attributes are typically gleaned from primary biometric data, are classifiable in pre-defined human understandable categories, and can be extracted in an automated manner.}'' The authors further note that soft biometrics can be used beyond recognition purposes \cite{dantcheva_what_2016}.
Using the notion of ``\textit{characteristics}'' in Article 4.14 instead of trait or feature, further reinforces that the goal is ``\textit{unique identification}'', as a characteristic is a unique/distinctive trait.

The GDPR provides no interpretation for ``specific technical processing'' although it is a pivotal moment for understanding and managing potential risks.
A biometric system performing recognition has two phases: enrolment and recognition. In \cite{bygrave_article_2020} the authors note that such processes consist of multiple steps, listing seven higher level points which can be considered as specific technical processing. 
Point (a) in their list states: ``\textit{acquiring a reference measure of one or more physical, physiological or behavioural characteristics of a person (often termed ‘enrolment’)}''. 
Enrolment can be further broken down to technical steps \cite{jain_50_2016}. These steps can be different in other modes of operation. For recognition these include:
(1) capture the biometric trait of the individual (2) create a biometric sample, (3) feature extraction (4) creation of templates (5) storing templates as biometric reference data.
Collecting and storing facial images and fingerprints (biometric samples) are not considered as biometric data or processing of biometric data under the GDPR (unlike in the ISO standard) as reflected by Recital 51 \cite{GDPR} and also pointed out by Kindt \cite{kindt_having_2018}. This means that ``specific technical processing'' only starts at the step of (3) feature extraction, but even that and step (4),(5) is tied to ``purpose of use'' whether it can produce data legally considered as biometric, as the result must allow recognition.
This view of specific technical processing where biometric systems extracts unique traits, to be used solely during the enrolment or the recognition phase, completely neglects other potential applications of biometric systems.

\section{Classification as a mode of operation}
WP29 notes that biometric systems in addition to recognition can be also used for other purposes: \textit{``The categorisation/segregation of an individual by a biometric system is typically the process of establishing whether the biometric data of an individual belongs to a group with some predefined characteristic in order to take a specific action [...] }''.
While unique biometric traits are most suitable for recognition purposes, both unique and soft measurable traits are classifiable. For example a fingerprint can be used in all three discussed modes of operation, it can identify, verify or classify an individual. An example for the latter can be gender classification from fingerprint ridge count and fingertip size \cite{gnanasivam_gender_2019}.

In multi-modal systems soft traits can increase performance, and even using only soft traits for identification is possible \cite{dantcheva_bag_2011,bourlai_classification_2020}, but their most common use is for classification purposes \cite{kindt_privacy_2013,van_der_hof_normative_2011, narang_gender_2016}.
A general system performing classification, instead of ``enrolment and recongition'' performs ``training and classification'', usually based on machine learning (ML) algorithms. 
There are several approaches for training these classifiers, the discussion of which is outside the scope of this paper.
Training can be performed on a data set that is independent from the data subject. Potential biases introduced during training such as sampling bias, over-fitting data, or temporal bias are among the few problems that could have a negative impact on rights and freedoms of the individuals. Under certain circumstances user data can be reused for training and correction, but for most individuals the first interaction with the biometric system will happen in the classification phase.

The primary classification tasks in ML are: binary classification where data is segregated into two sets, like classifying faces into female/male, thereby gendering biometric data on the way \cite{agapito_privacy_2015}. In multi-class classification there are more than two pre-established sets. Multi-label classification is typical in video surveillance scenarios, where age, gender, hair style, etc. can be determined from a single image. In imbalanced classification the number of cases for each class is not equally distributed. 

The phase of classification contains: (1) capture of the biometric traits, (2) sample creation, (3) feature extraction. (4) Create biometric probe(s) (5) labelling by trained ML classifier (6) produce classification decision(s).
For certain applications there can be a step, in conjunction with the capture process, which is linking the measured signals to a certain activity, like presenting a stimulus, and observing the performed action of the individual. Such processing is not subject to Article 9 but can have far reaching consequences in profiling applications, and combined with certain modalities can contribute to increased risks.

\section{Discussion}

Reflecting on the example from the introduction, presenting a stimulus and measuring the emotional reactions of individuals for the purpose of determining their emotional state, is not under the protection of Article 9. We present the following non-exhaustive list of risks in connection to classification purposes considering both unique and soft traits.

\textbf{Risk 1:} Data that researchers and other definitions would already call biometric, needs to satisfy condition 1. and 2. from Fig. 1, to reach this category in the GDPR. If condition 1. (specific technical processing) is related to classification steps/purposes (i.e. either the training or the classification phase), resulting data in most cases won't qualify as biometric.


\textbf{Risk 2:} 
Due to the data minimisation principle, if the intended purpose is not identification  the extracted feature should not identify the data subject. Still if recognition is not the purpose, and strong anonymization or obfuscation would render the data unusable, the controller is not obliged to anonymise and could as a result handle biometric data that confirms identity but doesn't get the protection of Article 9.

\textbf{Risk 3:} Classifying biometric traits of individuals, while they are identified by non-bio\-me\-tric means like a username, is not prohibited by Article 9, as usernames are not biometric data and the purpose of processing of the biometric data in questions is not identification. This can lead to individual profiling using physical, physiological or behavioural traits and characteristics of the data subject.

While determining if someone has a moustache might not appear to be a high-risk scenario, McStay points out far reaching consequences for the use of emotional surveillance in AI applications \cite{mcstay_emotional_2020}.
Psychological or cognitive biometrics relies on the measurement of cognitive or emotional states of an individual linked to certain activities \cite{obaidat_biometric-based_2019}. These states of mind are deducted from physical or behavioral actions/reactions, or bio-signals such as the electroencephalogram (EEG), electrocardiogram (ECG), or electrodermal response (EDR) of the individual in response to the presentation of a certain stimulus, e.g., viewing an image portraying a memorable event. 
These can be unique like EEG suitable for biometric recognition, or soft (non-unique) like measuring the emotional reaction to certain videos from watch time, pupil dilation or a survey. 

\textbf{Risk 4:} Technological advancements will allow such psychological-based techniques to be more accurate, continuously present and immersive. This can change the impact or the level of risks such techniques can impose to the rights and freedoms of individuals. Therefore, consequences and harms may change as well as likelihood. 

Psychological states such as emotions are measurable through several modalities. 
If such data is not considered to be biometric by the GDPR, another way to be eligible for special protection is through other special categories of data from Article 9, like racial or ethnic origin, data concerning health etc. As Kindt points out the GDPR doesn't confirm or reject this interpretation \cite{kindt_having_2018}. For certain physiological characteristics like cardiac signals it is more obvious that they qualify as health data, but this can't be generalised.

\textbf{Risk 5:} If high quality data  about physical, physiological or behavioural traits and characteristics are collected for classification purposes in large amounts, the controller might acquire data that reveals information about health, ethnicity, sexual orientation or other special categories of data, thus accumulating a large amount of sensitive personal data that will need special attention.  


\textbf{Risk 6:} Personal data about physical, physiological, behavioural traits and characteristics, whether or not they can satisfy the conditions to be regarded as biometric data, their processing for classification purposes escapes the protection of Article 9.

This implies that personal data about physical, physiological, behavioural traits can even undergo ``specific technical processing'' without being protected by Article 9, which further elevates potential risks and harms to the data subject.


\textbf{Risk 7:} According to Recital 26 \cite{GDPR}, when data is anonymized and the data controller is certain the data subject is not possible to be re-identified by any means and can demonstrate that, the processing of such anonymous data, including statistical or research purposes is not subject to the GDPR. Such data can still be used to evaluate even sensitive aspects of groups of people, leading to group profiling using sensitive information, the collection of which escapes Article 9 for classification purposes, in the context of biometric data. For data to remain unlikable, a sufficient anonymity set of similar data must be present, which is not always guaranteed \cite{pfitzmann_anonymity_2007}. 

The definition of classification is not synonymous with profiling but it can ``evaluate certain personal aspects'' (Article 4.4 GDPR), and can have similar consequences to profiling which is a serious privacy risk discussed. 
If such processing enters the category of Article 22 automated decision making including profiling, paragraph 1 states: ``\textit{The data subject shall have the right not to be subject to a decision based solely on automated processing, including profiling, which produces legal effects concerning him or her or similarly significantly affects him or her.}'' 
However, this is a weaker protection compared to Article 9. It is not prohibited by default and it is not obvious how data subjects can exercise their rights. 
Although the prohibitions of Article 9 can be lifted if a legal basis is fulfilled, it provides a stronger protection. On the contrary, regarding classification or individual profiling, processing might happen regardless or even in spite of the will of the data subject, as Article 22.2.a states paragraph 1 shall not apply: ``if is necessary for entering into, or performance of, a contract between the data subject and a data controller''. For example a legitimate legal bases can be direct marketing.

\textbf{Risk 8:} If Article 22.2(a) is fulfilled it causes the data subject to lose the right not to be subject to such processing operations, and explicit consent is not required anymore, even for classification purposes using unique or soft biometric modalities.

While Article 22, Recital 70 and 71 \cite{GDPR} reinforce that the data subject shall have the right to object to direct marketing, automated decision making and profiling, the interest of the controller can override the interests or the fundamental rights and freedoms of the data subjects, as pointed out by Veale et al. \cite{veale_when_2018} and reflected in Recital 69 \cite{GDPR}. Paragraph 4 of Article 22 states the ``Decisions referred to in paragraph 2 shall not be based on special categories of personal data referred to in Article 9(1)'', but as we established certain biometric modalities and modes of operations will escape that protection. 

\section{Conclusion}
In this paper we have shown high-risks impacting data subject rights in connection with processing their physical, physiological, behavioural traits and characteristics for classification purposes. 
Unfortunately, as our problem representation showed, certain types of data are regarded as biometric by researchers or in standards but not by the GDPR. Even for data which qualifies as biometric, but used for purposes other than identification, the GDPR gives no more protection than general obligations. Processing certain types of soft biometrics including but not limited to ones from psychological-based techniques for classification, can present threats that rival those posed by the processing of unique traits for identification purposes. These risks and issues we brought forward in the discussion are further aiming to assist in the risk assessment step of data protection impact assessment, which we specifically recommend to use for biometric applications. Technological advancements can change risks associated with the processing of personal data related to physical, physiological, or behavioural traits. Therefore, a more inclusive systematisation, and a more technologically neutral definition by the legislator would be beneficial.

\bibliographystyle{IEEEtran}
\bibliography{main}

\begin{thebibliography}{10}
\providecommand{\url}[1]{#1}
\csname url@samestyle\endcsname
\providecommand{\newblock}{\relax}
\providecommand{\bibinfo}[2]{#2}
\providecommand{\BIBentrySTDinterwordspacing}{\spaceskip=0pt\relax}
\providecommand{\BIBentryALTinterwordstretchfactor}{4}
\providecommand{\BIBentryALTinterwordspacing}{\spaceskip=\fontdimen2\font plus
\BIBentryALTinterwordstretchfactor\fontdimen3\font minus
  \fontdimen4\font\relax}
\providecommand{\BIBforeignlanguage}[2]{{%
\expandafter\ifx\csname l@#1\endcsname\relax
\typeout{** WARNING: IEEEtran.bst: No hyphenation pattern has been}%
\typeout{** loaded for the language `#1'. Using the pattern for}%
\typeout{** the default language instead.}%
\else
\language=\csname l@#1\endcsname
\fi
#2}}
\providecommand{\BIBdecl}{\relax}
\BIBdecl

\bibitem{GDPR}
\BIBentryALTinterwordspacing
{European Parliament and Council}, ``\BIBforeignlanguage{English}{{Regulation
  ({EU}) 2016/679 of the European Parliament and of the Council of 27 April
  2016 on the protection of natural persons with regard to the processing of
  personal data and on the free movement of such data, and repealing Directive
  95/46/{EC} (General Data Protection Regulation) (Text with {EEA}
  relevance)}},'' 2016. [Online]. Available:
  \url{http://data.europa.eu/eli/reg/2016/679/oj/eng}
\BIBentrySTDinterwordspacing

\bibitem{article_29_working_party_wp193_2012}
{Article 29 Working Party}, ``{WP193: Opinion 3/2012 on developments in
  biometric technologies},'' 2012.

\bibitem{meints_biometric_2008}
M.~{Meints}, ``Biometric {Systems} and {Data} {Protection} {Legislation} in
  {Germany},'' in \emph{2008 {International} {Conference} on {Intelligent}
  {Information} {Hiding} and {Multimedia} {Signal} {Processing}}, 2008, pp.
  1088--1093.

\bibitem{jain_50_2016}
A.~K. Jain, K.~Nandakumar, and A.~Ross, ``\BIBforeignlanguage{en}{50 years of
  biometric research: {Accomplishments}, challenges, and opportunities},''
  \emph{\BIBforeignlanguage{en}{Pattern Recognition Letters}}, vol.~79, pp.
  80--105, Aug. 2016.

\bibitem{4815263}
J.~{Ortega-Garcia}, ``The multiscenario multienvironment biosecure multimodal
  database (bmdb),'' \emph{IEEE Transactions on Pattern Analysis and Machine
  Intelligence}, vol.~32, no.~6, pp. 1097--1111, June 2010.

\bibitem{kindt_having_2018}
E.~J. Kindt, ``Having yes, using no? about the new legal regime for biometric
  data,'' \emph{Computer Law \& Security Review}, vol.~34, no.~3, pp. 523--538,
  2018-06-01.

\bibitem{kramer_experimental_2014}
A.~D.~I. Kramer, J.~E. Guillory, and J.~T. Hancock,
  ``\BIBforeignlanguage{en}{Experimental evidence of massive-scale emotional
  contagion through social networks},''
  \emph{\BIBforeignlanguage{en}{Proceedings of the National Academy of
  Sciences}}, no.~24, 2014.

\bibitem{bacchi_why_2012}
C.~Bacchi, ``Why {Study} {Problematizations}? {Making} {Politics} {Visible},''
  \emph{Open Journal of Political Science}, 2012.

\bibitem{schatten_towards_2009}
\BIBentryALTinterwordspacing
M.~Schatten, ``Towards a {General} {Definition} of {Biometric} {Systems},''
  2009. [Online]. Available: \url{https://arxiv.org/abs/0909.2365}
\BIBentrySTDinterwordspacing

\bibitem{bygrave_article_2020}
\BIBentryALTinterwordspacing
L.~A. Bygrave and L.~Tosoni, ``\BIBforeignlanguage{en}{Article 4(14).
  {Biometric} data},'' in \emph{\BIBforeignlanguage{en}{The {EU} {General}
  {Data} {Protection} {Regulation} ({GDPR})}}.\hskip 1em plus 0.5em minus
  0.4em\relax Oxford University Press, Feb. 2020. [Online]. Available:
  \url{https://doi.org/10.1093/oso/9780198826491.001.0001}
\BIBentrySTDinterwordspacing

\bibitem{jasserand_avoiding_2015}
\BIBentryALTinterwordspacing
C.~A. Jasserand, ``\BIBforeignlanguage{en}{Avoiding terminological confusion
  between the notions of ‘biometrics’ and ‘biometric data’: an
  investigation into the meanings of the terms from a {European} data
  protection and a scientific perspective},''
  \emph{\BIBforeignlanguage{en}{International Data Privacy Law}}, p. ipv020,
  Sep. 2015. [Online]. Available:
  \url{https://academic.oup.com/idpl/article-lookup/doi/10.1093/idpl/ipv020}
\BIBentrySTDinterwordspacing

\bibitem{jasserand_legal_2016}
C.~Jasserand, ``Legal {Nature} of {Biometric} {Data}: {From} ‘{Generic}’
  {Personal} {Data} to {Sensitive} {Data},'' \emph{European Data Protection Law
  Review}, vol.~2, 2016.

\bibitem{dantcheva_what_2016}
A.~Dantcheva, P.~Elia, and A.~Ross, ``What {Else} {Does} {Your} {Biometric}
  {Data} {Reveal}? {A} {Survey} on {Soft} {Biometrics},'' \emph{IEEE
  Transactions on Information Forensics and Security}, vol.~11, no.~3, pp.
  441--467, Mar. 2016.

\bibitem{gnanasivam_gender_2019}
P.~Gnanasivam and R.~Vijayarajan, ``\BIBforeignlanguage{en}{Gender
  classification from fingerprint ridge count and fingertip size using optimal
  score assignment},'' \emph{\BIBforeignlanguage{en}{Complex \& Intelligent
  Systems}}, vol.~5, no.~3, pp. 343--352, Oct. 2019.

\bibitem{dantcheva_bag_2011}
A.~Dantcheva, C.~Velardo, A.~D’Angelo, and J.-L. Dugelay,
  ``\BIBforeignlanguage{en}{Bag of soft biometrics for person
  identification},'' \emph{\BIBforeignlanguage{en}{Multimedia Tools and
  Applications}}, vol.~51, no.~2, pp. 739--777, 2011.

\bibitem{bourlai_classification_2020}
N.~Narang and T.~Bourlai, ``\BIBforeignlanguage{en}{Classification of {Soft}
  {Biometric} {Traits} {When} {Matching} {Near}-{Infrared} {Long}-{Range}
  {Face} {Images} {Against} {Their} {Visible} {Counterparts}},'' in
  \emph{\BIBforeignlanguage{en}{Securing {Social} {Identity} in {Mobile}
  {Platforms}}}, T.~Bourlai, P.~Karampelas, and V.~M. Patel, Eds.\hskip 1em
  plus 0.5em minus 0.4em\relax Cham: Springer International Publishing, 2020.

\bibitem{kindt_privacy_2013}
E.~J. Kindt, \emph{Privacy and Data Protection Issues of Biometric
  Applications: A Comparative Legal Analysis}.\hskip 1em plus 0.5em minus
  0.4em\relax Springer Netherlands, 2013.

\bibitem{van_der_hof_normative_2011}
I.~van~der Ploeg, ``\BIBforeignlanguage{en}{Normative {Assumptions} in
  {Biometrics}: {On} {Bodily} {Differences} and {Automated}
  {Classifications}},'' in \emph{\BIBforeignlanguage{en}{Innovating
  {Government}}}, S.~van~der Hof and M.~M. Groothuis, Eds.\hskip 1em plus 0.5em
  minus 0.4em\relax The Hague, The Netherlands: T. M. C. Asser Press, 2011,
  vol.~20, pp. 29--40, series Title: Information Technology and Law Series.

\bibitem{narang_gender_2016}
N.~Narang and T.~Bourlai, ``Gender and ethnicity classification using deep
  learning in heterogeneous face recognition,'' in \emph{2016 International
  Conference on Biometrics}, 2016.

\bibitem{agapito_privacy_2015}
\BIBentryALTinterwordspacing
A.~Othman and A.~Ross, ``\BIBforeignlanguage{en}{Privacy of {Facial} {Soft}
  {Biometrics}: {Suppressing} {Gender} {But} {Retaining} {Identity}},'' in
  \emph{\BIBforeignlanguage{en}{Computer {Vision} - {ECCV} 2014 {Workshops}}},
  L.~Agapito, M.~M. Bronstein, and C.~Rother, Eds.\hskip 1em plus 0.5em minus
  0.4em\relax Cham: Springer International Publishing, 2015, vol. 8926, pp.
  682--696, series Title: Lecture Notes in Computer Science. [Online].
  Available: \url{http://link.springer.com/10.1007/978-3-319-16181-5\_52}
\BIBentrySTDinterwordspacing

\bibitem{mcstay_emotional_2020}
A.~McStay, ``Emotional {AI}, soft biometrics and the surveillance of emotional
  life: {An} unusual consensus on privacy,'' \emph{Big Data \& Society, SAGE
  Publications Ltd}, vol.~7, no.~1, p. 2053951720904386, Jan. 2020.

\bibitem{obaidat_biometric-based_2019}
M.~S. Obaidat, I.~Traore, and I.~Woungang, Eds., \emph{Biometric-Based Physical
  and Cybersecurity Systems}.\hskip 1em plus 0.5em minus 0.4em\relax Springer
  International Publishing, 2019.

\bibitem{pfitzmann_anonymity_2007}
A.~Pfitzmann, T.~Dresden, M.~Hansen, and U.~Kiel,
  ``\BIBforeignlanguage{en}{Anonymity, {Unlinkability}, {Undetectability},
  {Unobservability}, {Pseudonymity}, and {Identity} {Management} – {A}
  {Consolidated} {Proposal} for {Terminology}},'' p.~84, 2007.

\bibitem{veale_when_2018}
M.~Veale, R.~Binns, and J.~Ausloos, ``\BIBforeignlanguage{en}{When data
  protection by design and data subject rights clash},''
  \emph{\BIBforeignlanguage{en}{International Data Privacy Law}}, vol.~8,
  no.~2, pp. 105--122, 2018.

\end{thebibliography}

\end{document}